\documentclass[twocolumn,twoside,aps,superscriptaddress,pre,showpacs]{revtex4}
\usepackage{graphicx}
\usepackage{dcolumn}
\usepackage{amssymb,amsmath}
\usepackage[dvips]{color}
\def\v{{\mbox{\boldmath$v$}}}
\def\r{{\mbox{\boldmath$r$}}}
\def\e{{\mbox{\boldmath$e$}}}
\def\n{{\mbox{\boldmath$n$}}}

\newcommand{\kB}{k}
\newcommand{\Txy}{T_{\!h\!o\!r}}
\newcommand{\fz}{F}
\newcommand{\fpp}{G}
\newcommand{\PP}{N}
\newcommand{\nnu}{\vartheta}
\newcommand{\rad}{r_{\!p}}
\newcommand{\length}{L}
\newcommand{\erf}{\mathop{\mathrm{erf}}}

\begin{document}
\bibliographystyle{apsrev}

\title{Temperature Anisotropy in a Driven Granular Gas}

\author{Devaraj van der Meer}
\affiliation{Department of Applied Physics and J.M. Burgers Centre
for Fluid Dynamics, University of Twente, P.O. Box 217, 7500 AE
Enschede, The Netherlands}
\author{Peter Reimann}
\affiliation{Department of Physics, University of Bielefeld, 33615
Bielefeld, Germany}



\begin{abstract}
When smooth granular material is fluidized by vertically shaking a
container, we find that the temperature in the direction of energy
input always exceeds the temperature in the other directions. An
 analytical model is presented which shows how the anisotropy
 can be traced back to the inelasticity of the interparticle collisions and
 the collisions with the wall. The model compares very well with
 molecular dynamics simulations. It is concluded 
 that any non-isotropic driving of a granular gas in a steady state
 {\it necessarily} causes anisotropy of the granular temperature.
\end{abstract}

\maketitle

As compared to an ordinary, molecular gas, the hallmark of a
granular gas is its permanent dissipation of energy due to
inelastic collisions. Whereas an isolated molecular gas will
sustain its motion for an infinite amount of time, the only true
equilibrium state of granular matter is the one where it is at
rest. Hence a steady supply of energy is required to keep a
granular gas alive, giving rise to prototypical non-equilibrium
systems with many striking phenomena \cite{goldhirsch03,weele03}.
The one addressed in this Letter is the crucial 
temperature anisotropy within a granular gas. 
It is observed to be a significant effect in both numerical
simulations
\cite{mcnamara98,barrat02,herbst04,brey98cub,krouskop03,krouskop03b}
and experiments \cite{wildman00,yang02,kudrolli01,kudrolli03}.
Although it has been studied in the context of a random
restitution coefficient model \cite{barrat01,barrat03}, a
theoretical explanation is still lacking. Here we provide such an
explanation, in which for analytical convenience we will restrict
ourselves to a dilute granular gas, fluidized by vertically
vibrating a container.

So what causes the anisotropy? We approach this question by a
theoretical model in combination with event driven molecular
dynamics (MD) simulations, and show that the anisotropy results
from the following characteristic feature of such a gas: The {\it
distribution} of energy from the vibrating bottom towards the
horizontal directions occurs through the very same mechanism that
also constitutes one of the major sources of energy {\it
dissipation}, i.e., the collisions between the particles. This
result carries over to any granular gas with a non-isotropic
energy source.

\begin{figure}
\includegraphics*[scale=0.28]{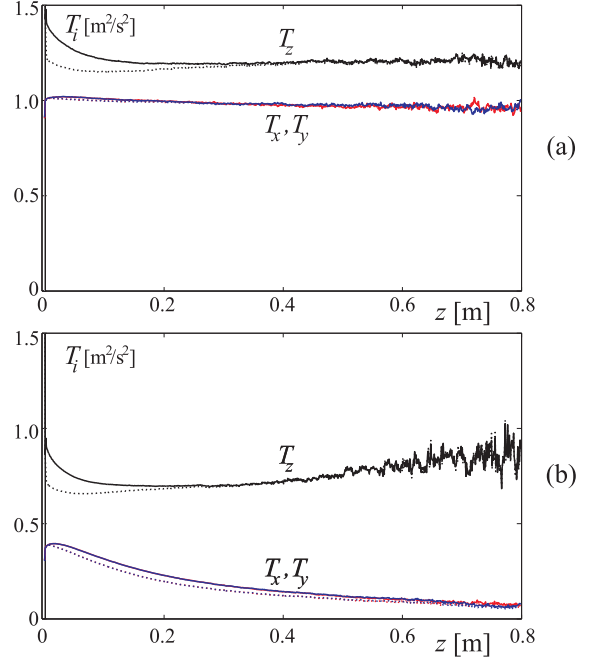}
\caption{\small Solid lines: Temperature components (\ref{eq01})
from MD simulations of the driven granular gas model as specified
in the main text with sinusoidal driving and parameters $f=72.648$
Hz, $a=1.00$ mm, $\PP =50$, $\rad = 1.18$ mm, and $\length =2.265$
cm. (a): $e_p=0.9$ and $e_w=1$ (reflective sidewalls). (b):
$e_p=0.95$ and $e_w=0.95$ (dissipative sidewalls). Dashed lines:
Counting only particles that move downward, i.e. restricting the
average in (\ref{eq01}) to negative $v_z$-values. At large $z$ the
statistics deteriorates since particles rarely reach such heights.
The curves $T_x(z)$ and $T_y(z)$ practically coincide.}
\label{Fig1}
\end{figure}

The setup we will consider in our present work consists of a
granular gas in a container with a square-shaped bottom of side
length $\length$ in the $x$-$y$-plane and infinitely high,
vertical side-walls. Gravity acts with $g=9.81$ m/s$^2$ and the
gas is fluidized by vertical vibrations of the bottom about $z=0$
with amplitude $a$ and frequency $f$ -- typically of triangular
(piecewise linear, symmetric) or sinusoidal shape. The gas
consists of $\PP$ identical hard spheres with radius $\rad$ and
mass $m$. We restrict ourselves to the case
\cite{mcnamara98,sela98,brey98,baldassari01,brey01,eggers99,soto99}
that only normal restitution \cite{goldhirsch03,weele03}
contributes to the dissipative processes, with restitution
coefficients $e_p$ for particle-particle and $e_w$ for
particle-wall collisions, while collisions with the vibrating
bottom are taken to be perfectly elastic.

After initial transients have died out, we expect (and observe) a
stationary probability distribution $\rho(\r,\v)$ of the particle
positions $\r:=(x,y,z)$ and velocities $\v:=(v_x,v_y,v_z)$. The
quantities of central interest are the temperature components
\begin{equation}
T_i(z):=\frac{m}{\kB}\,\langle v_i^2 \rangle \label{eq01}
\end{equation}
where $i\in\{x,y,z\}$ and $\kB$ is Boltzmann's constant. This
$T_i(z)$ is directly proportional to the average kinetic energy of
the particles in a horizontal layer at height $z$ in either of the
three spatial directions $i$.

In evaluating our MD simulations we replace ensemble averages in
(\ref{eq01}) by time averages (justified by ergodicity) and work
in units with $m=1$ and $\kB=1$. A representative result is
depicted in Fig.~\ref{Fig1}: As expected for symmetry reasons, the
horizontal temperature components $T_x(z)$ and $T_y(z)$ are
practically indistinguishable. In contrast, the vertical
temperature component $T_z(z)$ is significantly larger than
$T_x(z)$ and $T_y(z)$. For perfectly elastic particle-wall
collisions (Fig.~\ref{Fig1}a) the $z$-dependence of the
temperature components is very weak except for a region directly
above the bottom. There, the energy-input by the driving leads to
increased upward particle velocities, as shown by the dashed lines
in (Fig.~\ref{Fig1}a). For inelastic particle-wall collisions
(Fig.~\ref{Fig1}b) the $z$-dependence of the temperature
components is more pronounced. Yet, in both cases the differences
between vertical and horizontal temperatures are more important
than the $z$-dependences.

Our theoretical analysis of the observed temperature differences
starts with the well-known conservation laws of energy and
momentum for a dilute granular gas, derived from Boltzmann's
equation \cite{sela98,brey98,baldassari01}. For a stationary
system, they read, in terms of the local heat flux
$\mathbf{J}(\r)$ and stress (or pressure) tensor $\mathbf{P}(\r)$:
\begin{eqnarray}
\nabla \cdot \mathbf{J}(\r) = I(\r) \ , \ \ \nabla \cdot
\mathbf{P}(\r) = n(\r) \mathbf{f}(\r) \ .  \label{eq1}
\end{eqnarray}
Here, $I(\r)$ is the local energy dissipation rate per unit
volume, $n(\r)$ is the local particle density, and the force
$\mathbf{f}(\r)= -m g \mathbf{e}_z$. Integrating the first
equation (\ref{eq1}) over the container volume $V$ we obtain with
Gauss' theorem that the energy dissipation rate due to
particle-particle collisions $Q_{pp} := \int_V I(\r) \,
d\mathbf{r}$ must be equal to the total flux of energy through the
boundaries. The latter can be decomposed in the influx $Q_{in} :=
\length ^2\, J_z(0)$ of energy through the vibrating bottom of
area $\length ^2$ and the energy dissipation rate $Q_w$ due to
particle-wall collisions. This gives
\begin{equation}
Q_{in} =  Q_{pp} + Q_w . \label{eq4}
\end{equation}

\begin{figure}
\includegraphics*[scale=0.32]{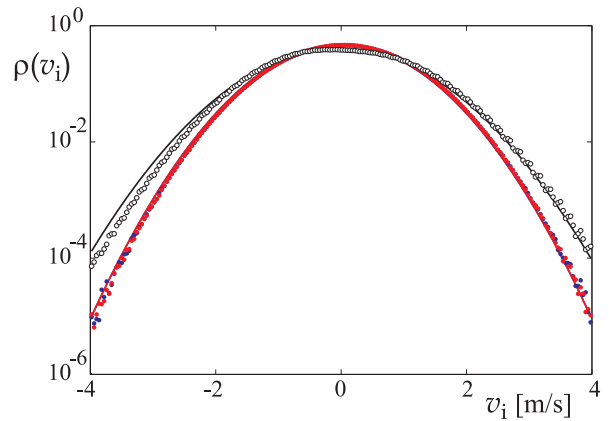}
\caption{\small Horizontal (solid red and blue dots) and vertical
(open, black dots) velocity distributions of particles in a layer
between $z=0$ and $z=5$ mm obtained by MD simulations for the same
system as in Fig. \ref{Fig1}a but with $f=30$ Hz and $\length
=4.8$ cm. The solid lines are Gaussian fits \cite{f2}.}
\label{Fig2}
\end{figure}

Crucial to the present model is that the temperature components
$T_i$ defined in (\ref{eq01}) are treated separately. To our
knowledge, all existing theories for driven granular gases in a
steady state without net flow of material are based upon the
assumption of an isotropic temperature, and many of them also on
isotropic stress \cite{sela98,brey98,baldassari01,brey01,f1}.

The temperature components $T_i$ are related to the diagonal
elements of the stress tensor by a generalized ideal gas law
$P_{ii} = n \kB T_i$ \cite{sela98,brey98,baldassari01}. Motivated
by our MD simulations, we assume that each temperature component
$T_i$ is approximately constant within the entire container volume
(see also \cite{kumaran98,eggers99}). Because of symmetry the
stress tensor is diagonal and $T_x=T_y=:\Txy$, with which the
second equation (\ref{eq1}) can now be readily integrated to yield
\begin{equation}
n(\r) = \frac{\PP mg}{\length ^2 \kB T_z}
\exp \left\{- \frac{mg z}{\kB T_z}\right\} \ . \label{eq3}
\end{equation}
Furthermore, as exemplified by Fig.~\ref{Fig2}, our MD simulations
show that the particle velocity components can be assumed as
Gaussian distributed in very good approximation for a wide range
of parameter values \cite{olafsen99,f2}. All together, we thus
arrive at the following approximative distribution function for
particle position and velocity:
\begin{eqnarray}
    \rho(\r,\v) = \frac{n(\r)\ m^{3/2}}
    {\sqrt{(2\pi\kB )^3\Txy^2 T_z }}
    \exp\!\left\{\!-\frac{m(v_x^2+x_y^2)}{2\kB\Txy}
    -\frac{mv_z^2}{2\kB T_z}\!\right\} \ .
    \label{eq6}
\end{eqnarray}

The next main idea is to determine the two unknowns $T_z$ and
$\Txy$ in (\ref{eq6}) by means of two energy balance relations.
The first of them is (\ref{eq4}). To obtain the second, we observe
that particle-particle collisions not only cause energy
dissipation but also a transfer of kinetic energy from the
horizontal direction into the vertical direction, and vice versa.
In the steady state the net effect must be an average loss of
kinetic energy per time unit $Q_z$ in the vertical direction which
is exactly balanced by the incoming energy flux $Q_{in}$ through
the vibrating bottom:
\begin{eqnarray}
Q_{in} =  Q_{z}  \ .
\label{eq5}
\end{eqnarray}

The remaining, rather technical task is to explicitly determine
all the energy fluxes $Q$ appearing in (\ref{eq4}) and (\ref{eq5})
with the help of the approximation (\ref{eq6}).
In order to evaluate $Q_z$, we first note that
the change of kinetic energy in the vertical direction
in a single particle-particle collision is
\begin{equation}
q_z = \sum_{j=1}^2 \frac{m}{2}
\left( (\v_j \cdot \e_z)^2 - (\v^*_j \cdot \e_z)^2 \right) \ ,
\label{eq7}
\end{equation}
where $\v_j$ and $\v^*_j$ are the velocities of the two colliding
particles ($j=1,2$) before and after the collision, respectively.
Due to our assumption that only normal restitution contributes to
the dissipative processes, we have $\v^*_j = \v_j + (-1)^j
(1+e_p)[(\v_1-\v_2)\cdot\n]\n/2$, where $\n$ is the collision
normal vector. To determine $Q_z$ one essentially has to introduce
this result for $\v^*_j$ into (\ref{eq7}) and then average
according to (\ref{eq6}). More precisely, first $q_z$ in
(\ref{eq7}) is multiplied by the collisional volume per unit time
$(1/2) \pi (2\rad)^2 |\v_2-\v_1|\delta(|\r_1 - \r_2|-2\rad)$,
where the factor $1/2$ arises since collisions only can happen if
$(\v_2-\v_1)\cdot\n<0$. Next we multiply with
$(1/2)\rho(\r_1,\v_1) \rho(\r_2,\v_2)$ according to (\ref{eq6})
and integrate over all $\v_j$ and $\r_j$ (within the container
volume). The factor $1/2$ is needed since every collision appears
twice in the above considerations. This gives, after a substantial
amount of algebra:
\begin{eqnarray}
Q_z \!\! &=& \!\! (1 + e_p) \, (\PP \rad/\length )^2 g \sqrt{\pi m \kB T_z} \,  \fz (\Txy/T_z)
\label{eq8}
\\
\fz (\nnu) \!\! &:=& \!\! \int\limits_0^\infty \!ds\!
\int\limits_0^\infty \! dt\, \sqrt{s + t}\, \frac{(8-4e_p)s -
(1+e_p) t}{6\, \nnu \sqrt{s}\, \exp \{s+t/\nnu \}} \ . \nonumber
\end{eqnarray}
A similar averaging of the total energy loss in a single
particle-particle collision
$q_{pp} = (m/2) \sum_{j=1}^2 ( \v_j^2 - (\v^*_j)^2 )$ yields
\begin{eqnarray}
Q_{pp}\!\! &=& \!\! (1 - e_p^2) \, (\PP \rad/\length )^2 g \sqrt{\pi m \kB T_z} \,  \fpp (\Txy/T_z)
\label{eq10}
\\
\fpp (\nnu) \!\! &:=& \!\! \int\limits_0^\infty \!ds\!
\int\limits_0^\infty \!dt \, \frac{[s + t]^{3/2}}{\nnu \sqrt{s}\,
\exp \{s+t/\nnu \}} \ . \nonumber
\end{eqnarray}
In the same spirit one can evaluate the total dissipation rate
due to particle-wall collisions $Q_w$ with the result
\begin{equation}
Q_{w} = 4\,(1-e_w^2)\, \PP \, (\kB \Txy)^{3/2} / (\length  \sqrt{2\pi m})
\label{eq12}
\end{equation}
Finally, a somewhat lengthy but straightforward calculation yields
the following expression for the energy input rate at the perfectly elastic,
vibrating bottom of the container:
\begin{equation}
Q_{in} = 2\, \PP \, g \, m^{3/2} \, u^2 \psi(\gamma)/\sqrt{\pi \kB T_z}
\label{eq13}
\end{equation}
where $\gamma := \sqrt{2 \kB T_z / m u^2}$, $u=:\pi\sqrt2\, a f$
for sinusoidal driving of the bottom, and $u:=4\,a f$ for
triangular driving. In both cases, for large $\gamma$ (which is
the typical situation in the dilute systems under study) the
function $\psi(\gamma)$ approaches unity \cite{f3}.

\begin{figure}
\includegraphics*[scale=0.27]{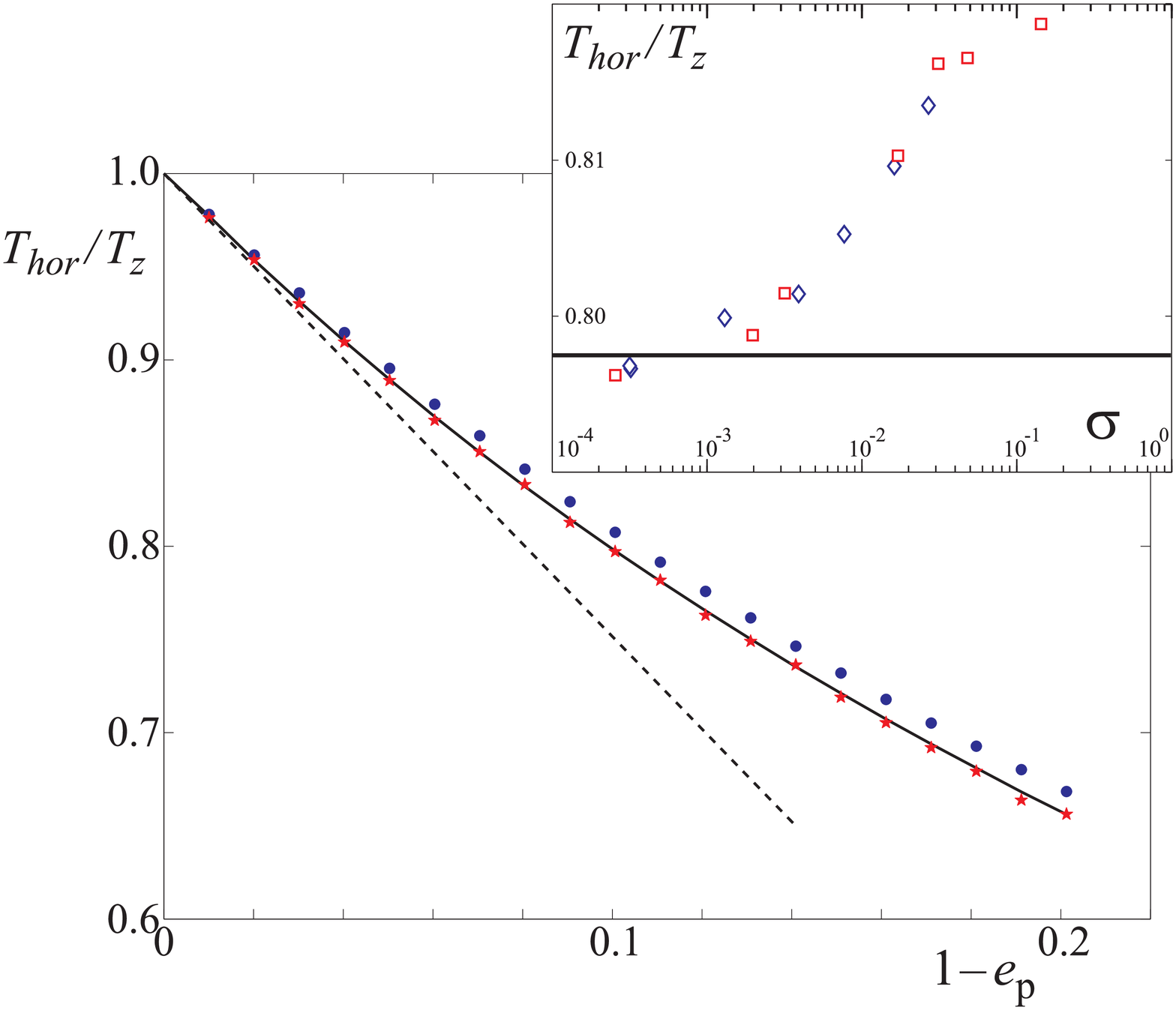}
\caption{\small Ratio of horizontal and vertical temperatures
versus particle-particle restitution coefficient. Solid line:
Theoretical prediction (\ref{eq11}). Dashed line: Theoretical
asymptotic behavior (\ref{eq11a}). Symbols: MD simulations for the
same system as in Fig. \ref{Fig1}a with $\PP =20$ (\textcolor{red}{$\star$}
) and $\PP =50$ (\textcolor{blue}{$\bullet$}
) particles. For $e_p\not = 0.9$, a different choice for the
frequency $f$ such that the $T_z$ was (approximately) the same for
all $e_p$. Inset: Same, but for fixed $e_p=0.9$ and variable local
solid fraction $\sigma := (4/3) \pi \rad^3 \max[n({\bf r})]$,
either by varying particle number $\PP$
(\textcolor{blue}{$\Diamond$}, at $r=1.18$ mm) or radius $\rad$
(\textcolor{red}{$\Box$}, at $N=50$). 
}
\label{Fig3}
\end{figure}

In the absence of wall dissipation ($e_w=1$), Eqs.~(\ref{eq4}),
(\ref{eq5}), (\ref{eq8})-(\ref{eq13}) imply
\begin{equation}
\fz (\Txy/T_z) = (1 - e_p)\fpp(\Txy/T_z) \ , \ \ [e_w=1] \ .
\label{eq11}
\end{equation}
Closer inspection shows that for any $0 \leq e_p<1$ a unique
solution $0 \leq \Txy/T_z<1$ of (\ref{eq11}) exists.
In particular, for small $1-e_p$ one finds the leading order
asymptotics
\begin{equation}
\Txy/T_z = 1 - (5/2)\, (1-e_p) \ , \ \ [e_w=1] \ .
\label{eq11a}
\end{equation}
Thus for $e_p<1$ and perfectly reflecting walls, the model
predicts that the horizontal temperature $\Txy$ is {\it always}
smaller than the vertical temperature $T_z$. Moreover, the ratio
$\Txy/T_z$ solely depends on $e_p$ but not on any details of
driving shape and strength, particle density, particle size, or
compartment geometry. The comparison with MD simulations in
Fig.~\ref{Fig3} is excellent. The inset of Fig.~\ref{Fig3} shows
that the remaining deviations can largely be attributed to finite
size effects.

\begin{figure}
\includegraphics*[scale=0.26]{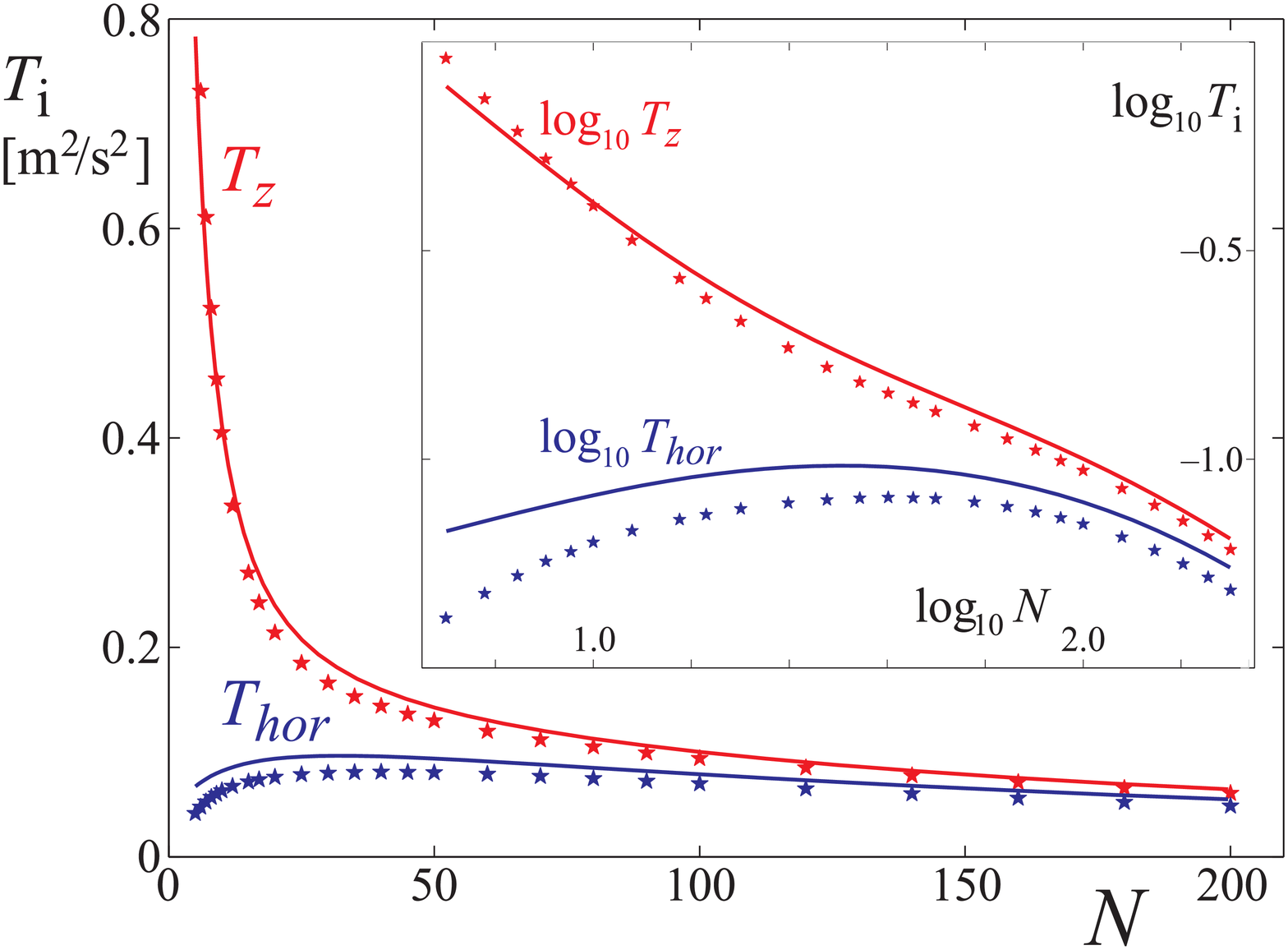}
\caption{\small Horizontal (blue) and vertical (red) temperatures
versus particle number. Lines: Theoretical predictions from
(\ref{eq4}), (\ref{eq5}), (\ref{eq8})-(\ref{eq13}) for triangular
driving and parameters $f=30$ Hz, $a=1.00$ mm, $\rad = 1.18$ mm,
$\length =4.8$ cm, $e_p=0.95$ and $e_w=0.95$. Symbols: MD
simulations. Inset: Same data in a log-log plot. } \label{Fig4}
\end{figure}

In the general case one obtains two transcendental, algebraic
equations for the two unknowns $\Txy$ and $T_z$ by introducing
(\ref{eq8})-(\ref{eq13}) into (\ref{eq4}) and (\ref{eq5}). While
existence and uniqueness of solutions can still be demonstrated
analytically, their quantitative determination is only possible
numerically. An example is depicted in Fig.~\ref{Fig4}, comparing
very well with MD simulations. As expected, dissipative walls tend
to reduce the horizontal temperature $\Txy$ since they add a
source of dissipation for the horizontal kinetic energy. If we
increase the particle number, the gas becomes more dense, and the
number of particle-particle collisions will increase much faster
than the number of particle-wall collisions. Therefore, $\Txy$
will first {\it increase}, in sharp contrast to the overall
temperature $[T_z+2\Txy]/3$ which must decrease with increasing
particle density. Eventually, particle-particle collisions will
dominate the system, and $\Txy/T_z$ will asymptotically tend to
the value it would have with reflecting walls.

In conclusion, we have numerically observed large differences
between the vertical and horizontal temperatures in vertically
driven granular gases subjected to gravity. We introduced a
theoretical model based on an approximative Maxwell-Boltzmann
distribution with anisotropic but homogeneous temperature
(\ref{eq6}), justified by our MD simulations. Both for reflecting
and dissipative walls of the container we find that the
theoretical model gives good quantitative agreement with the
simulations.

The difference of the horizontal and vertical temperatures from
the isotropic value is a significant correction, at least of the
same order as those resulting from taking into account the
non-constancy of the temperature and density profiles, or from
embedding Chapman-Enskog corrections to the velocity distributions
into the theoretical framework \cite{sela98,brey98}.

At the root of the temperature anisotropy lies the fact that the
{\it transfer} of kinetic energy between different spatial
directions and its {\it dissipation} arise out of the same
mechanism: the collisions between the particles. Thus, the
anisotropy of the temperature is a {\it necessary} consequence of
the anisotropy of the driving. The present work indicates that one
may obtain an improved hydrodynamic description for dilute
granular gases by starting out
from an anisotropic velocity distribution 
(instead of an isotropic one) in deriving hydrodynamics equations
from Boltzmann's equation.

The basic concept of the model applies to any situation in which
the energy-input in a granular gas is anisotropic, always
predicting a higher kinetic energy content in the main direction
of energy-input.

\acknowledgments We thank Detlef Lohse and Ko van der Weele for
many useful suggestions and discussions. This work is part of the
research program of the Stichting FOM, which is financially
supported by NWO. P.R. acknowledges support by the Deutsche
Forschungsgemeinschaft under SFB 613 and RE 1344/3-1, and by the
ESF-program STOCHDYN.


\begin{thebibliography}{99}

\bibitem{goldhirsch03} I. Goldhirsch, {\it Annu. Rev. Fluid Mech.} {\bf 35},
267 (2003).

\bibitem{weele03}
K.~van~der Weele, R.~Mikkelsen, D.~van~der Meer, and D.~Lohse,
 in \textit{The Physics of Granular Media}, D.~Wolf and H.~Hinrichsen (ed.), vol. in press
  (Wiley-VCH, 2004).

\bibitem{mcnamara98} S. McNamara and S. Luding, {\it Phys. Rev. E} {\bf 58},
813 (1998).

\bibitem{brey98cub} J. Brey and D. Cubero, 
{\it Phys. Rev. E} {\bf 57}, 2019 (1998).

\bibitem{barrat02} A. Barrat and  E. Trizac,
{\it Phys. Rev. E} {\bf 66}, 051303 (2002).

\bibitem{krouskop03} P.E. Krouskop and J. Talbot, 
cond-mat/0303263 (preprint, 2003).

\bibitem{krouskop03b} P.E. Krouskop and J. Talbot, 
cond-mat/0312530 (preprint, 2003).

\bibitem{herbst04} O. Herbst, P.
M{\"u}ller, M. Otto, and A.
Zippelius, cond-mat/0402104 (preprint, 2004). 

\bibitem{yang02} X. Yang, C. Huan, D. Candela, R.W.
Mair, and R.L. Walsworth, {\it Phys. Rev. Lett.} {\bf 88}, 044301
(2002).

\bibitem{wildman00} R.D. Wildman and J.M. Huntley, {\it Powder
Technology} {\bf 113}, 14 (2000). 

\bibitem{kudrolli01} D.L. Blair and A. Kudrolli, {\it Phys. Rev.
E} {\bf 64}, 050301 (2001).

\bibitem{kudrolli03} D.L. Blair and A. Kudrolli, {\it Phys. Rev.
E} {\bf 67}, 041301 (2003).

\bibitem{barrat01} A. Barrat,  E. Trizac, and J. N. Fuchs, 
{\it Eur. Phys. J. E} {\bf 5}, 161 (2001).

\bibitem{barrat03} A. Barrat and  E. Trizac,
{\it Eur. Phys. J. E} {\bf 11}, 99 (2003).

\bibitem{sela98} N. Sela and I. Goldhirsch, {\it J. Fluid
Mech.} {\bf 361}, 41 (1998).

\bibitem{brey98} J.J. Brey, J.W. Dufty, C.S. Kim, and
A. Santos, {\it Phys. Rev. E} {\bf 58}, 4638 (1998).

\bibitem{baldassari01} A. Baldassarri, U. Marini Bettolo Marconi,
A. Puglisi, and A. Vulpiani, {\it Phys. Rev. E} {\bf 64}, 011301
(2001).

\bibitem{brey01} J.J. Brey, M.J. Ruiz-Montero, and F. Moreno, {\it
Phys. Rev. E} {\bf 63}, 061305 (2001).

\bibitem{eggers99} J. Eggers, {\it Phys. Rev. Lett.} {\bf 83}, 5322
(1999).

\bibitem{soto99}
R. Soto, M. Mareschal, and D. Risso, Phys. Rev. Lett. {\bf 83},
5003 (1999).

\bibitem{kumaran98} V. Kumaran, {\it Phys. Rev. E} {\bf 57}, 5660
(1998).

\bibitem{olafsen99} J.S. Olafsen and J.S. Urbach, {\it Phys. Rev.
E} {\bf 60}, R2468 (1999).

\bibitem{campbell86} C.S. Campbell and A. Gong, {\it J. Fluid
Mech.} {\bf 164}, 107 (1986).

\bibitem{goldhirsch96} I. Goldhirsch and N. Sela, {\it Phys. Rev.
E} {\bf 54}, 4458 (1996).

\bibitem{jenkins88} J.T. Jenkins and M.W. Richman, {\it J. Fluid
Mech.} {\bf 192}, 313 (1988).

\bibitem{kumaran99} V. Kumaran, {\it Phys. Rev. Lett.} {\bf 82},
3248 (1999).

\bibitem{brey00} J.J. Brey, M.J. Ruiz-Montero, and F. Moreno, {\it
Phys. Rev. E} {\bf 62}, 5339 (2000).

\bibitem{f1}
Only in the context of the normal-stress differences observed in
steady plane Couette flow of granular material \cite{campbell86},
an anisotropic stress tensor \cite{goldhirsch96} or
Maxwell-Boltzmann velocity distribution \cite{jenkins88} has been
used.

\bibitem{f2}
In \cite{brey00,kumaran99}, large deviations from Gaussian
velocity distributions, especially near the vibrating bottom, have
been reported. Our simulations show that they are rooted in the
{\em discontinuous}, saw-tooth shape of the vibrations considered
in \cite{brey00,kumaran99}. This is the main reason that we focus
on {\em continuous} shapes of the driving in the present work.

\bibitem{f3}
For triangular driving we have derived the exact result
$\psi(\gamma)=(1/4)[2 + 3\exp(-1/\gamma^2) - 6 \exp(-4/\gamma^2) +
3 \exp(-9/\gamma^2)]
+(\sqrt{\pi}/8\gamma)[(6+3\gamma^2)\erf(1/\gamma) -
(24+2\gamma^2)\erf(2/\gamma) + (18+\gamma^2)\erf(3/\gamma)]$. For
sinusoidal driving we were only able to show that $\psi(\gamma)\to
1$ for $\gamma\to\infty$.

\end{thebibliography}
\end{document}